\documentclass[aps,amsmath,amssymb,reprint,superscriptaddress]{revtex4-1}
\usepackage{graphicx,dcolumn,bm}
\usepackage[colorlinks,linkcolor=blue,anchorcolor=blue,citecolor=blue]{hyperref}
\begin{document}

\title{Generation of ten kilotesla longitudinal magnetic fields in ultraintense laser-solenoid target interactions}

\author{K. D. Xiao}
\affiliation{Center for Applied Physics and Technology, HEDPS, and School of Physics, Peking University, Beijing 100871, People's Republic of China}
\author{C. T. Zhou}
\email[]{Electronic mail: zcangtao@sztu.edu.cn}
\affiliation{Center for Applied Physics and Technology, HEDPS, and School of Physics, Peking University, Beijing 100871, People's Republic of China}
\affiliation{Center for Advanced Material Diagnostic Technology, Shenzhen Technology University, Shenzhen 518118, People's Republic of China}
\affiliation{College of Optoelectronic Engineering, Shenzhen University, Shenzhen 518060,  People's Republic of China}
\affiliation{Institute of Applied Physics and Computational Mathematics, Beijing 100094, People's Republic of China}
\author{H. Zhang}
\affiliation{Institute of Applied Physics and Computational Mathematics, Beijing 100094, People's Republic of China}
\author{T. W. Huang}
\affiliation{Center for Advanced Material Diagnostic Technology, Shenzhen Technology University, Shenzhen 518118, People's Republic of China}
\affiliation{College of Optoelectronic Engineering, Shenzhen University, Shenzhen 518060,  People's Republic of China}
\author{R. Li}
\affiliation{Center for Applied Physics and Technology, HEDPS, and School of Physics, Peking University, Beijing 100871, People's Republic of China}
\author{B. Qiao}
\affiliation{Center for Applied Physics and Technology, HEDPS, and School of Physics, Peking University, Beijing 100871, People's Republic of China}
\author{J. M. Cao}
\affiliation{Center for Advanced Material Diagnostic Technology, Shenzhen Technology University, Shenzhen 518118, People's Republic of China}
\affiliation{College of Optoelectronic Engineering, Shenzhen University, Shenzhen 518060,  People's Republic of China}
\author{T. X. Cai}
\affiliation{Center for Advanced Material Diagnostic Technology, Shenzhen Technology University, Shenzhen 518118, People's Republic of China}
\author{S. C. Ruan}
\affiliation{Center for Advanced Material Diagnostic Technology, Shenzhen Technology University, Shenzhen 518118, People's Republic of China}
\affiliation{College of Optoelectronic Engineering, Shenzhen University, Shenzhen 518060,  People's Republic of China}
\author{X. T. He}
\affiliation{Center for Applied Physics and Technology, HEDPS, and School of Physics, Peking University, Beijing 100871, People's Republic of China}
\affiliation{Center for Advanced Material Diagnostic Technology, Shenzhen Technology University, Shenzhen 518118, People's Republic of China}
\affiliation{Institute of Applied Physics and Computational Mathematics, Beijing 100094, People's Republic of China}

\date{\today}
\begin{abstract}
Production of the huge longitudinal magnetic fields by using an ultraintense laser pulse irradiating a solenoid target is considered. Through three-dimensional particle-in-cell simulations, it is shown that the longitudinal magnetic field up to ten kilotesla can be observed in the ultraintense laser-solenoid target interactions. The finding is associated with both fast and return electron currents in the solenoid target. The huge longitudinal magnetic field is of interest for a number of important applications, which include controlling the divergence of laser-driven energetic particles for medical treatment, fast-ignition in inertial fusion, etc., as an example, the well focused and confined directional electron beams are realized by using the solenoid target.


\end{abstract}
\maketitle

Huge magnetic field generation in laboratory has attracted more and
more attention due to its wide applications, including plasma
physics\cite{LiangE}, laboratory astrophysics\cite{CiardiA}, atomic
and nuclear physics\cite{MurdinB}, and material
sciences\cite{Kojima}, etc. In laser driven inertial confinement
fusion, the huge magnetic field, which is generated by the magnetic
flux compression of the seed field, increases the plasma temperature
of the hot spot and confines the alpha particles to the burn
region\cite{changpy}.

At present, the strongest continuous magnetic field generated by the
hybrid magnet is around $45\ \rm{T}$, and the strongest pulsed
magnetic field generated by the non-destructive electromagnet is
around $100\ \rm{T}$\cite{HayesI}. Besides of the magnet, high power
lasers have the potential to generate huge magnetic fields of
extreme strengths. The magnetic field generated by a nanosecond
laser ablation on a capacitor-coil target is reported $1.5\ \rm{kT}$
at the GEKKO-XII laser facility\cite{FujiokaS}, $800\ \rm{T}$ at the
LULI pico 2000 laser facility\cite{SantosJ}, and $610\ \rm{T}$ at
the GEKKO-LFEX laser facility\cite{LawK}. The magnetic field $205\
\rm{T}$ is reported to be generated by the open-ended coil at the
SG-II laser facility\cite{ZhuB}. With the rapid development of the
laser technologies, the laser facilities with higher intensities
($10^{19}\sim10^{21}\rm{W/cm^2}$) and shorter durations (picoseconds
to femtoseconds) have been built
up\cite{PukhovA,DaidoH-rev,MacchiA}. When such an ultraintense laser
pulse interacts with plasmas, a large number of hot electrons are
produced by the laser ponderomotive acceleration. These hot
electrons generate currents at the plasma-vacuum interfaces with
magnitude $10^{16}\sim10^{18}\ \rm{A/m^2}\ $ \cite{Nakamura,ZhouC}.
The interface currents further induce a huge magnetic field with
strength $10^3\sim10^5\
\rm{T}$\cite{HainesM,BellA,SarriG,KarS,Tatarakis,Ramakrishna,PerezF},
which is much higher than that from the capacitor-coil with normal
discharging currents\cite{DaidoH}.
Thus, the huge magnetic field of extreme strength can be expected
when an ultraintense laser pulse interacts with a specially designed
target.

In this Letter, we propose to use an ultraintense laser pulse
irradiating solenoid target to generate the huge longitudinal
magnetic field, where such target design has also been applied on
proton acceleration\cite{KarS-mag}.
The solenoid target consists of a foil and a curved plasma wire. The
longitudinal static magnetic fields are induced by the surface currents
along the plasma wire  and
enhanced inside the solenoid, finally achieving almost uniform
spatial distribution. Three-dimensional particle-in-cell (PIC)
simulations are performed to study the generation of the
longitudinal magnetic field inside the solenoid target. As an
application, the magnetic field effects on guiding and focusing of
the hot electron beams are also studied. The PIC simulation
technique has been the main tool for studying laser-plasma
interaction and transport of electrons in hot, mildly dense plasma
near the relativistic critical density. For simulating overdense
plasmas one can use hybrid techniques \cite{bell-mag,
Robinson,camp,dav}, where high-temperature Spitzer resistivity and
low-temperature modifications are included. Here we shall use the
EPOCH code \cite{Arber}, which have been shown to yield good results
in similar laser-plasma interaction problems \cite{zam1, rid1, cer1}.

\begin{figure}
\includegraphics[width=6cm]{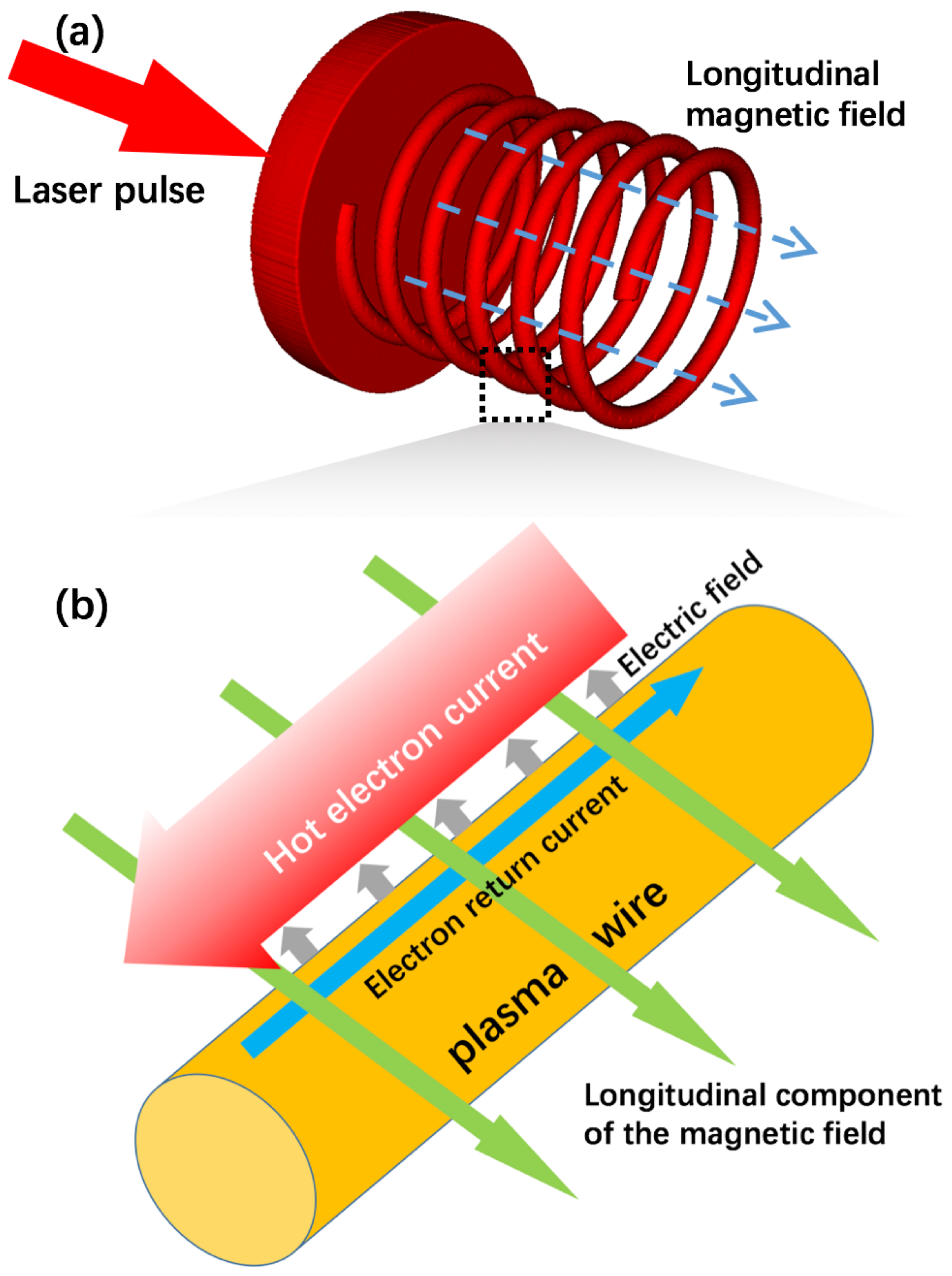}
\caption{\label{fig1} (Color online) (a) Configuration of the
solenoid target, where a curved plasma wire is attached at the
backside of a foil. The laser pulse irradiates from the left onto
the foil front surface. A backward longitudinal magnetic field is
generated inside the solenoid. (b) Schematic diagram of the
generation of the surface electric and magnetic fields. A small part
of the plasma wire is enlarged. The directions of the currents and
fields are denoted by arrows. See context for detail.}
\end{figure}

\begin{figure*}
\includegraphics[width=12cm]{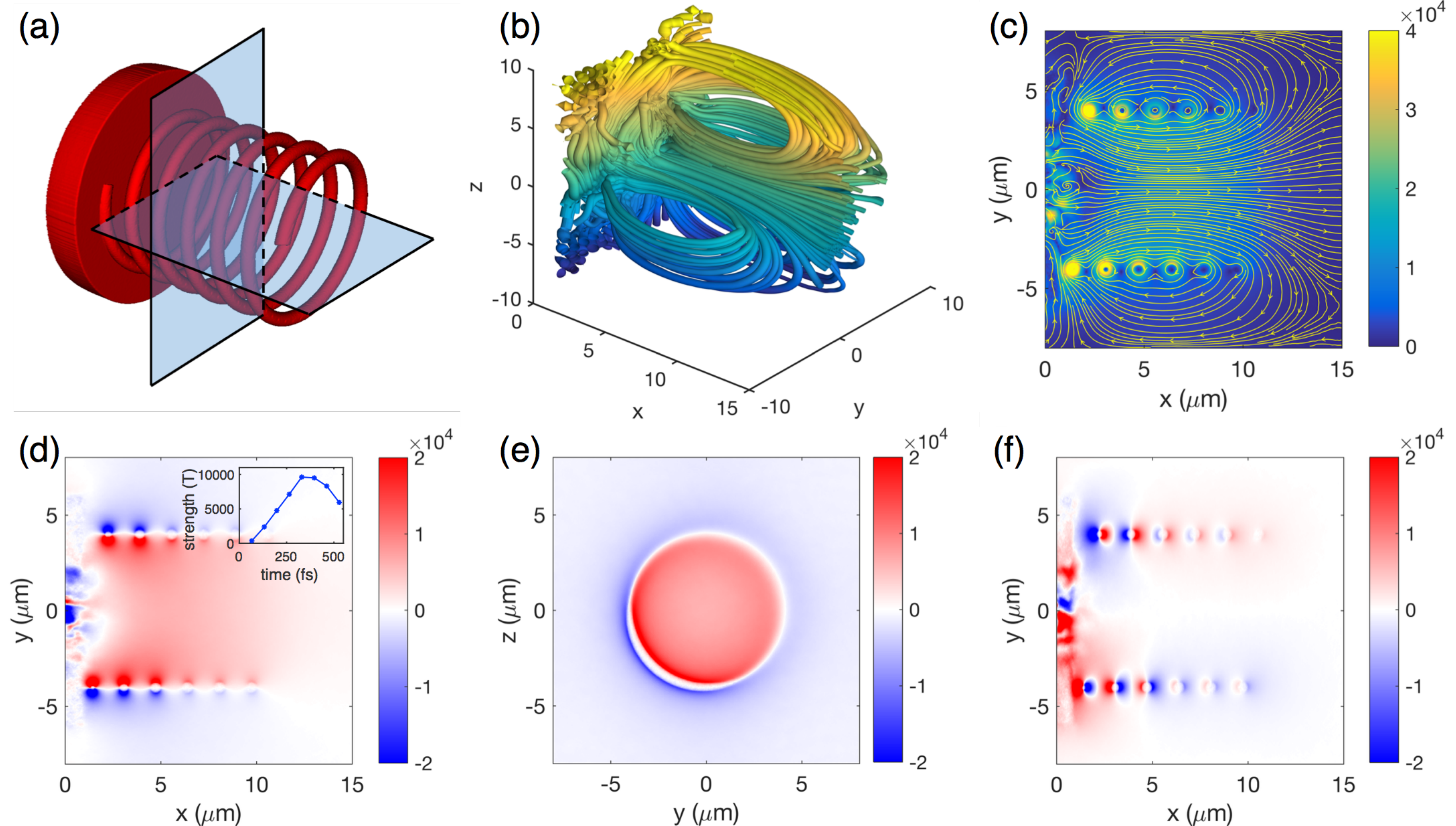}
\caption{\label{fig2} (Color online) (a) Positions of two
cross-section planes, which are $x$-$y$ plane ($z=0$ plane) and
$y$-$z$ plane ($x=5\ \rm{\mu m}$ plane). (b) Three-dimensional
distribution of the magnetic field lines . The lines facing forward are hidden to
look inside. The laser pulse is from
the left along the $x$-axis. (c) Magnetic force lines in the $x$-$y$
plane. (d, e) $B_x$ in the $x$-$y$ and $y$-$z$ planes. (f) $B_y$ in
the $x$-$y$ plane. The magnetic fields are taken at $t=330\ \rm{fs}$
and their unit is $\rm{T}$. Evolution of $B_x$ at (5, 0, 0)($\mu$m)
is shown in the insert of (d). }
\end{figure*}

The solenoid target consists of a foil and a curved plasma wire, as
shown in Fig. \ref{fig1}(a). The front of the plasma wire is
attached at the rear surface of the foil. The PIC simulation starts
from an ionized plasma composed of $\rm{Cu}^{2+}$ ions and
electrons. The electron number density of the target is
$n_{e0}=40n_c$, where
$n_c=m_e\epsilon_0\omega_0^2/e^2\approx1.11\times10^{27}\
\rm{m^{-3}}$ is the critical density. The initial temperatures are
$1\ \rm{keV}$ for the electrons and $170\ \rm{eV}$ for the ions. The
diameter and thickness of the foil are $12\ \rm{\mu m}$ and $1\
\rm{\mu m}$, respectively. The diameter and longitudinal length of
the solenoid are $d=8\ \rm{\mu m}$ and $h=10\ \rm{\mu m}$,
respectively. The number of coils contained in the solenoid is
$n=6$. The length of a single coil is $l=\sqrt{(\pi
d)^2+(h/n)^2}=25\ \rm{\mu m}$, and the total length of the solenoid
is $L=150\ \rm{\mu m}$. The diameter of the plasma wire is $0.6\
\rm{\mu m}$. A small scale preplasma is taken in front of the foil
with total length $5\ \rm{\mu m}$ and density profile
$n_e=n_{e0}\exp(x/\delta)$, where $\delta=0.5\ \rm{\mu m}$. The
laser pulse is $y$-polarized with intensity $2\times10^{20}\
\rm{W/cm^2}$ ($a_0\approx12.8$) and duration $100\ \rm{fs}$. The
laser wave length is $\lambda_0=1.06\ \rm{\mu m}$. The spatial
profile of the laser is Gaussian $a=a_0\exp(-r^2/\sigma^2)$ with the
spot radius $\sigma=3\ \rm{\mu m}$. The temporal profile of the
laser is flat top with raising and falling times of $1$ laser cycle.
The simulation box size at the $x\times y\times z$ directions are
$23\ \rm{\mu m}\times 16\ \rm{\mu m}\times 16\ \rm{\mu m}$,
respectively. The number of grids are $1143\times 795\times 795$,
respectively. The grid length equals to 1 plasma skin depth  $20\
\rm{nm}$. There is 6 macroparticles filled in each target cell. The
laser injection and open boundary conditions are taken at the
positive and negative $x$ boundaries, respectively. The periodic
boundary conditions are taken at the $y$ and $z$ boundaries.

When the ultraintense laser pulse irradiates the foil front surface,
a significant amount of hot electrons are created and accelerated
\cite{PukhovA}.
These hot electrons transit through the foil and some of them enter
the curved plasma wire and propagate along the latter as hot
electron currents.
Because of the Alfv\'en limit, cold return currents for balancing
the hot electrons are generated, as shown in Fig. \ref{fig1}(b). The
hot and return currents at the wire surface form a two-layer
structure, inducing a huge surface magnetic field. The hot electron
currents gradually expand into the vacuum, so that the spatial
distribution of the magnetic field spread towards the center of the
solenoid. The magnetic field strength at the center of the solenoid
is enhanced by the merging of the expanding fields. The spatial
distributions of the magnetic field and the force lines are shown in
Fig. \ref{fig2} at a time when the laser pulse is over. The magnetic
fields originating from the plasma wires have merged into a field
surrounding the solenoid. The magnetic field lines inside and
outside the solenoid are in opposite directions, as can be seen in
Figs. \ref{fig2}(b) and (c). The longitudinal magnetic field inside
the solenoid is almost uniformly distributed, as shown in Figs.
\ref{fig2}(d) and (e). The field strength is about $1\times10^4\
\rm{T}$ at the solenoid center and about $2\times10^4\ \rm{T}$ at
the plasma wire surface. The evolution of the magnetic field is
shown in the insert of Fig. \ref{fig2}(d). The magnetic field
strength increases almost linearly in the early time and reaches its
maximum value at $t=300\ \rm{fs}$. After that, the magnetic field
strength stays at the maximum value with slow decrease for about
$100\ \rm{fs}$, which is nearly the same as the laser pulse
duration. The magnetic field begins to vanish at $t=400\ \rm{fs}$,
when the hot electrons almost reach the solenoid end. At $t=500\
\rm{fs}$, the magnetic field strength inside the solenoid is about
$6000\ \rm{T}$.

\begin{figure}
\includegraphics[width=8.5cm]{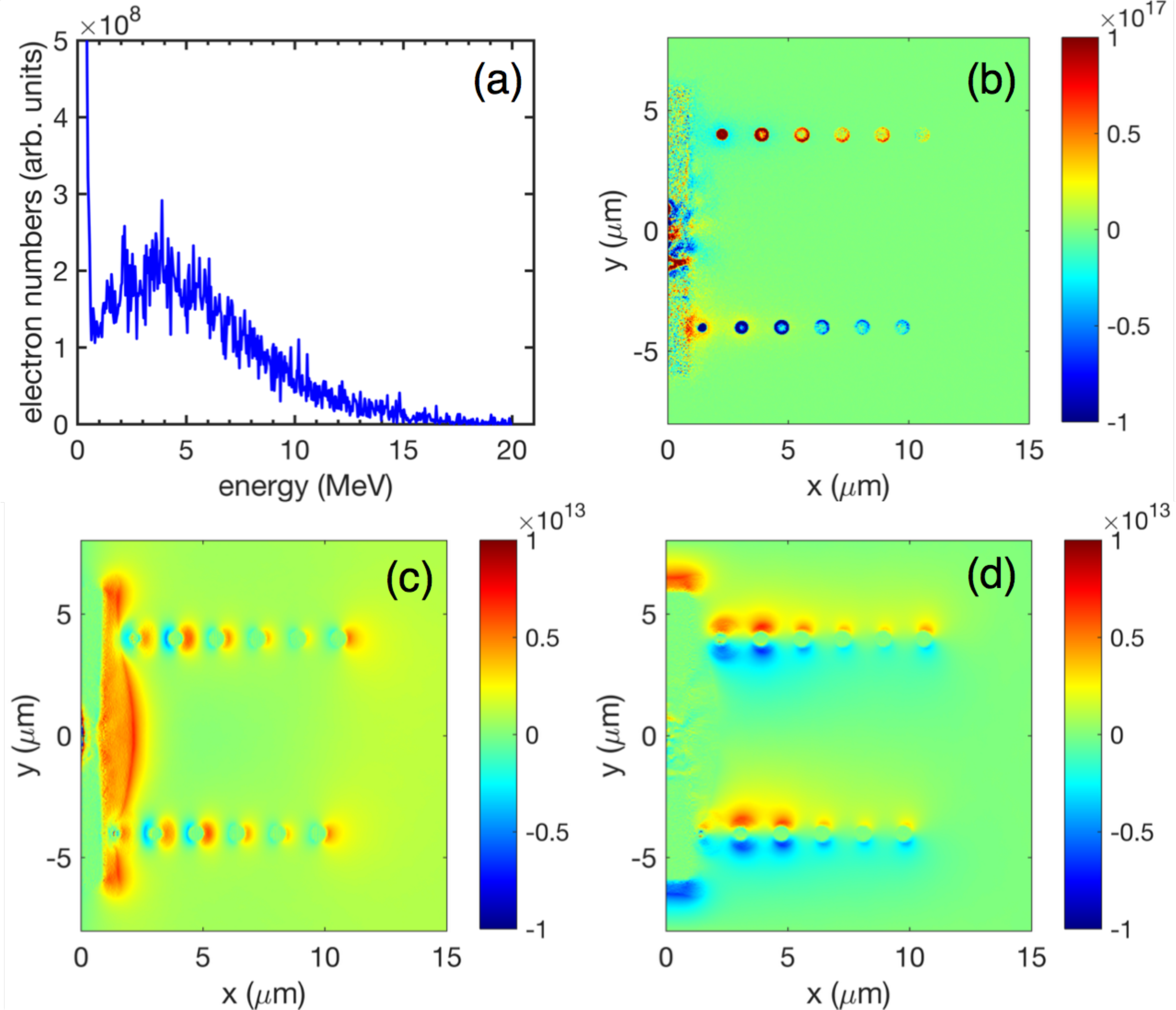}
\caption{(Color online) Distribution of the currents and electric
fields. (a) Spectrum for electrons in the plasma wire from $x=2\
\rm{\mu m}$ to $10\ \rm{\mu m}$. (b) Distribution of the electron
current $j_z$, (c) the longitudinal electric field $E_x$, and (d)
the transverse electric field $E_y$ at the cross-section plane
$x$-$y$ ($z=0$ plane). The spectrum and distributions are for $t=330\ \rm{fs}$.
The units for the electron current and electric fields are
$\rm{A/m^2}$ and $\rm{V/m}$, respectively.} \label{fig3}
\end{figure}

The hot electron propagation along the plasma wire is the key
mechanism for generation of the longitudinal magnetic field. The
energy distribution of wire electrons are shown in Fig. \ref{fig3}
(a). We note that most of the hot electrons moving along the plasma
wire have the energies around $5\ \rm{MeV}$, and the maximum
electron energy is about $20\ \rm{MeV}$. With the electron
propagation, a surface electric field surrounding the plasma wire is
induced, with maximum strength of about $7\times10^{12}\ \rm{V/m}$,
as shown in Figs. \ref{fig3}(c) and (d). The electric field can be
estimated by $E\approx T_h/eK$, where
$T_h\approx0.511[(1+I_{18}\lambda_0^2/2)^{1/2}-1] \rm{MeV}$ is the
hot-electron temperature and $K$ is the hot-electron spatial
extension. The distribution of the two-layer electron currents are
shown in Fig. \ref{fig3}(b) with the maximum current density of
about $1\times10^{17}\ \rm{A/m^2}$. From the electron current
distribution, the strength of the magnetic field at the wire surface
can be estimated by
$B_{\mathrm{MG}}\approx0.38n_{29}P_{\mathrm{TW}}^{-1}T_{h,511}R_{\mathrm{\mu
m}}T_{c,\mathrm{keV}}$\cite{bell-mag}, where $n_{29}$ is the
electron density in units of $10^{29}\ \rm{m^{-3}}$,
$P_{\mathrm{TW}}$ is the power of the hot electron beam in
$\rm{TW}$, $T_{h,511}$ is the hot electron temperature in units of
$511\ \rm{keV}$, $R_{\mathrm{\mu m}}$ is the hot electron beam
radius in $\rm{\mu m}$, and $T_{c,\mathrm{keV}}$ is the cold
background electron temperature in units of $\rm{keV}$. In the
simulations, the hot and cold electron temperatures are about $4.6\
\rm{MeV}$ and $250\ \rm{keV}$ respectively. The electron beam
intensity along the wire is about $P_e\sim\frac{c}{\Delta
L}\sum_{\Delta L}\epsilon_e\approx0.1\ \rm{TW}$, where $\epsilon_e$
is the electron energy for the electrons in wire with length $\Delta
L$. The beam electron density is nearly $8\times10^{27}\
\rm{m^{-3}}$ and the $R$ is taken as the plasma wire radius $0.3\
\rm{\mu m}$. From the above equation, the estimated magnetic field
strength at the wire surface is about $205\ \rm{MG}$ ($2\times10^4\
\rm{T}$), which agrees with our simulation result.

\begin{figure}
\includegraphics[width=8.5cm]{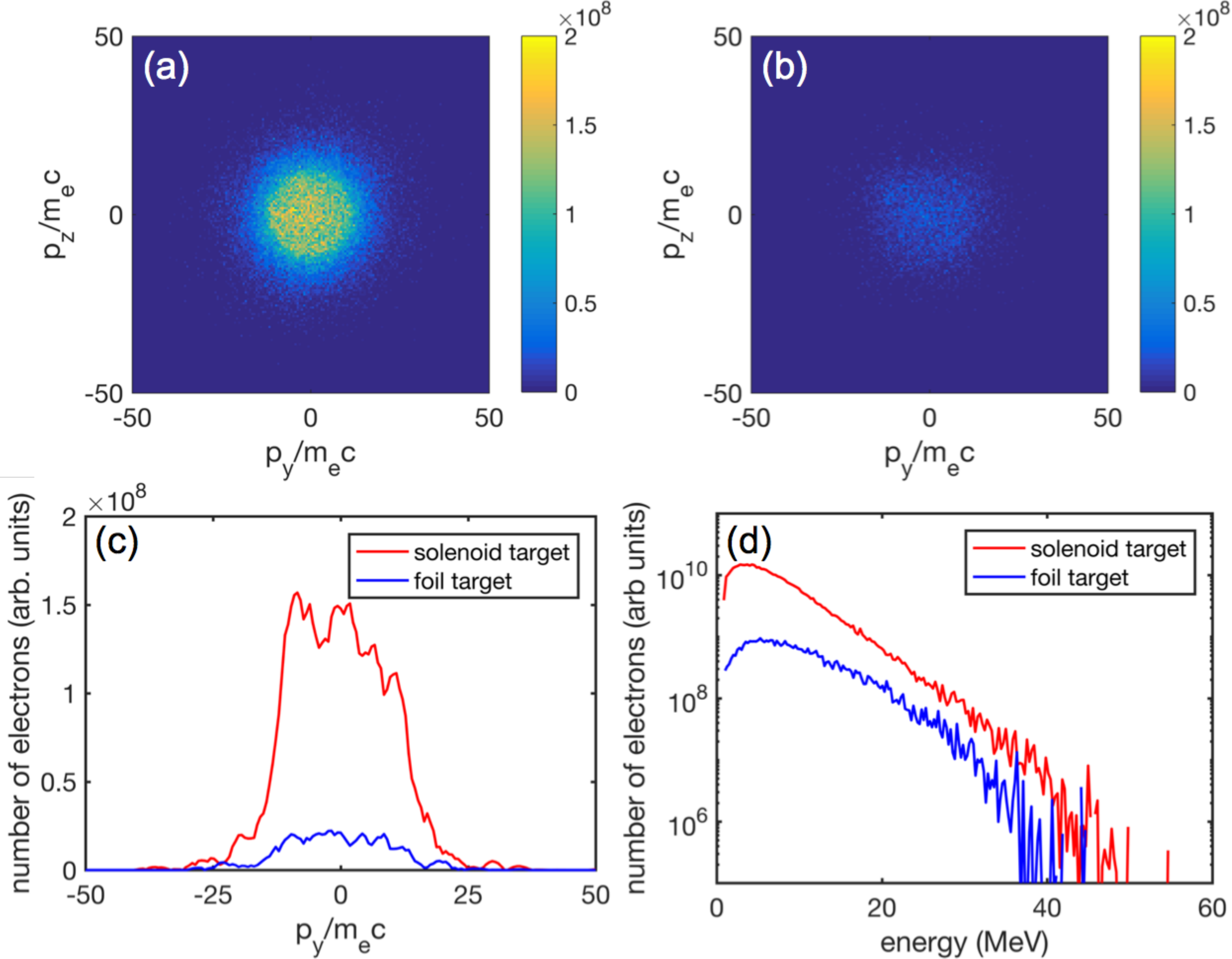}
\caption{(Color online) Time integral (from $t=0$ to $500\ \rm{fs}$)
of the electron numbers passing through a disc placed $8\ \rm{\mu
m}$ behind the foil. The disc radius is $5\ \rm{\mu m}$. (a)
Distribution of electron transverse momentum space for the electrons
with energies greater than $5\ \rm{MeV}$ for the solenoid target,
and (b) the bare foil target. (c) Corresponding transversal
distribution of the two momentum space. (d) Electron spectrum for
the time integral.} \label{fig4}
\end{figure} 

The relativistic electron beam (REB) has been widely used in laser
driven proton and radiation sources, isochoric heating of materials,
and warm dense matter production, etc.\cite{macchi, corde, perez,
norreys}. In previous work\cite{bailly, PerezF}, 
ones used externally longitudinal magnetic fields
to focus and guide the REB. Here we consider an integrated simulation to
simulateously generate both the collimating  magnetic fields and the REB using the solenoid
target.
In the solenoid, an electron in $x$-$y$ plane is affected by
the force $F_z=ev_yB_x-ev_xB_y$. After a time interval of $\Delta
t$, the $z$ component of the electron velocity is
$v_z=\frac{e}{m}B_xS_y-\frac{e}{m}B_yS_x$, where the magnetic field
strength is assumed constant during the small time interval, and
$S_x$ and $S_y$ are the electron displacements at the $x$ and $y$
directions, respectively. This transverse velocity $v_z$ leads to a
focusing force $F_y=-\frac{e^2}{m}B_x^2S_y+\frac{e^2}{m}B_xB_yS_x$.
For a given magnetic field $B_x=1\times10^4\ \rm{T}$, the maximum
energy of the focused electron is
$\mathcal{E}_{y}\sim\frac{e^2}{m}B_x^2S_y^2\approx17.6\ \rm{MeV}$
with focusing length $S_y=1\ \rm{\mu m}$, and
$\mathcal{E}_{y}\sim70.3\ \rm{MeV}$ with $S_y=2\ \rm{\mu m}$. To see
the REB focusing effect of the solenoid target, we have compared the
electron divergence of this target with that of a normal bare foil
target. The distribution of electrons passing through a time
integral  disc  placed $8\ \rm{\mu m}$
behind the front foil for the two cases is shown in Fig. \ref{fig4}.
The disc radius is $5\ \rm{\mu m}$. Figs. \ref{fig4}(a) and (b) show
that for the bare foil target the REB diverges behind the foil and
for the solenoid target it is confined and focused by the magnetic
field. A large number of hot electrons are trapped by the solenoid
magnetic field, forming a dense electron cloud at the center of the
momentum space, as shown in Fig. \ref{fig4}(c). The electron
spectrums for the two cases are shown in Fig. \ref{fig4}(d). It is
seen that the total hot electron number and the maximum electron
energy in the solenoid target is higher than that for the bare foil
target. Since the laser-foil interactions are same in the two cases,
the difference in the electron spectrum is related to the electron
motion behind the foil.
In the solenoid target case more electrons pass through the disc,
while in bare foil target case a large number of electrons are
diverged out of the disc. As a result, the REB generated by using
the solenoid target has large electron number and low divergence,
and is well confined and guided by the longitudinal magnetic field.

In summary, generation of huge longitudinal magnetic field can be
achieved by using a solenoid target. In our three-dimensional
particle-in-cell simulation, magnetic fields of strength nearly
$1\times10^4\ \rm{T}$ and duration nearly $500\ \rm{fs}$ is
generated by a laser pulse of intensity $2\times10^{20}\
\rm{W/cm^2}$ and duration $100\ \rm{fs}$. In this scheme, the laser
heated hot electrons flow along the solenoid plasma wire, generating
surface electric fields, and forming interface hot electron currents
and cold return currents. The two-layer currents further induce the
longitudinal magnetic field. The magnetic field is nearly uniform
distributed inside the solenoid, and the strength is far beyond the
traditional nanosecond laser-coil interaction methods. As an
application, the focusing and guiding of the hot electron beams by
the solenoid magnetic field is also studied.

This work is supported by the National Key Program for S$\&$T
Research and Development, Grant No.2016YFA0401100; the SSTDF, Grant
No. JCYJ20160308093947132; the National Natural Science Foundation
of China (NSFC), Grant Nos. 11575031, 11575298, 91230205, and
11705120. B. Q. acknowledges the support from Thousand Young Talents
Program of China.

{}

\end{document}